\documentclass[aip,amsmath,amssymb,reprint]{revtex4-1}

\usepackage{graphicx}
\usepackage{epstopdf}

\begin{document}

\title{Direct measurements of the magnetocaloric effect in pulsed magnetic fields: The example of the Heusler alloy Ni$_{50}$Mn$_{35}$In$_{15}$}

\author{M.~Ghorbani Zavareh}
\affiliation{Dresden High Magnetic Field Laboratory (HLD), Helmholtz-Zentrum Dresden-Rossendorf, 01328 Dresden, Germany.}
\affiliation{Institut f\"{u}r Festk\"{o}rperphysik, TU Dresden, 01062, Dresden, Germany.}
\author{C.~Salazar Mej\'{i}a}\email{Catalina.Salazar@cpfs.mpg.de}
\affiliation{Max Planck Institute for Chemical Physics of Solids, 01187 Dresden, Germany}
\author{A.~K.~Nayak}
\affiliation{Max Planck Institute for Chemical Physics of Solids, 01187 Dresden, Germany}
\author{Y.~Skourski}
\affiliation{Dresden High Magnetic Field Laboratory (HLD), Helmholtz-Zentrum Dresden-Rossendorf, 01328 Dresden, Germany.}
\author{J.~Wosnitza}
\affiliation{Dresden High Magnetic Field Laboratory (HLD), Helmholtz-Zentrum Dresden-Rossendorf, 01328 Dresden, Germany.}
\affiliation{Institut f\"{u}r Festk\"{o}rperphysik, TU Dresden, 01062, Dresden, Germany.}
\author{C.~Felser}
\affiliation{Max Planck Institute for Chemical Physics of Solids, 01187 Dresden, Germany}
\author{M.~Nicklas}
\affiliation{Max Planck Institute for Chemical Physics of Solids, 01187 Dresden, Germany}

\date{\today}

\begin{abstract}
We have studied the magnetocaloric effect (MCE) in the shape-memory Heusler alloy Ni$_{50}$Mn$_{35}$In$_{15}$ by direct measurements in pulsed magnetic fields up to 6 and 20~T. The results in 6~T are compared with data obtained from heat-capacity experiments. We find a saturation of the inverse MCE, related to the first-order martensitic transition, with a maximum adiabatic temperature change of $\Delta T_{ad} = -7$~K at 250~K and a conventional field-dependent MCE near the second-order ferromagnetic transition in the austenitic phase. The pulsed magnetic field data allow for an analysis of the temperature response of the sample to the magnetic field on a time scale of $\sim 10$ to 100~ms which is on the order of typical operation frequencies (10 to 100~Hz) of magnetocaloric cooling devices. Our results disclose that in shape-memory alloys the different contributions to the MCE and hysteresis effects around the martensitic transition have to be carefully considered for future cooling applications.
\end{abstract}


\maketitle

Recently, tremendous efforts have been made to find alternative technologies to replace the conventional gas compression/expansion technique for cooling applications.\cite{Zimm1998,Gschneidner2005} Higher efficiency cooling and environmental-friendly magnetic refrigeration based on the magnetocaloric effect (MCE) have initiated intensive research activity. Besides its practical application, MCE studies can give an extra insight in the properties of the magnetic phase transitions. Among the most promising materials, Heusler-type Ni-Mn-In magnetic shape-memory alloys are attractive candidates for both fundamental research and application. These alloys have been shown to exhibit diverse functional properties such as shape memory,\cite{Kainuma2006,Planes_JOPM_2009} magnetic superelasticity,\cite{Krenke_PRB_2007} magnetocaloric,\cite{Liu2012} and barocaloric effect,\cite{Manosa2010} which originate from magnetoelastic couplings. Ni$_{50}$Mn$_{35}$In$_{15}$ exhibits on cooling a paramagnetic to ferromagnetic transition around 315~K, followed by a first-order structural transition from a cubic high-temperature phase to a low-temperature monoclinic phase around 246~K, the so-called martensitic transition. A conventional MCE is observed around the ferromagnetic transition in the austenitic phase. Here, an increase in the sample temperature is observed upon application of a magnetic field in adiabatic conditions. Additionally, an inverse MCE is present around the martensitic first-order transition leading to a decrease in the sample temperature with increasing field. Due to the first-order character of the latter transition, hysteresis losses and fatigue are critical issues on the search of suitable materials for magnetocaloric cooling devices. So far, most MCE studies have been done by calculating the isothermal entropy change, $\Delta S_{iso}$, based on indirect methods, however, for first-order phase transitions additional considerations should be applied.\cite{Niemann2014} Direct measurements, besides giving the adiabatic temperature change, $\Delta T_{ad}$, which is one of the important parameters for magnetic refrigeration, are straightforward and closer to the actual process used in applications. Non-destructive pulsed-field facilities have typical pulse lengths of 10 to a few 100~ms, which match the targeted operation frequency of magnetic refrigerators, 10-100~Hz.\cite{Kuzmin2007} Besides extending the accessible magnetic field range and going beyond 50~T, the short pulse duration provides nearly adiabatic conditions during the measurement. Thus, direct MCE  measurements in pulsed magnetic fields provide the opportunity to investigate the dynamics of the MCE in a suitable frequency range. In this work, we study the MCE in a Ni$_{50}$Mn$_{35}$In$_{15}$ sample by direct magnetocaloric measurements in pulsed magnetic fields and by analyzing specific-heat data taken in static magnetic fields.

Polycrystalline ingots of Ni$_{50}$Mn$_{35}$In$_{15}$ were obtained by arc-melting stoichiometric amounts of the constituent elements under argon atmosphere. The ingots were remelted several times to assure a high homogeneity. Subsequently they were encapsulated in a quartz ampoule under argon atmosphere and annealed at 800$^\circ$C for 2~h and then quenched in ice water. The high quality of the samples was confirmed by powder X-ray diffraction. Magnetization and heat-capacity measurements were carried out in a physical property measurement system (Quantum Design). Direct measurements of $\Delta T_{ad}$  were performed in a home-built experimental setup in pulsed magnetic fields at the Dresden High Magnetic Field Laboratory (HLD). The temperature change of the sample was monitored by a Copper-Constantan thermocouple. The thermocouple was squeezed between two pieces of the sample to optimize the thermal contact and reduce the heat losses. The target temperature was approached with the help of a local heater. A resistive Cernox thermometer was utilized to determine the reference temperature.



\begin{figure}[t!]
\begin{center}
  \includegraphics[width=0.9\linewidth]{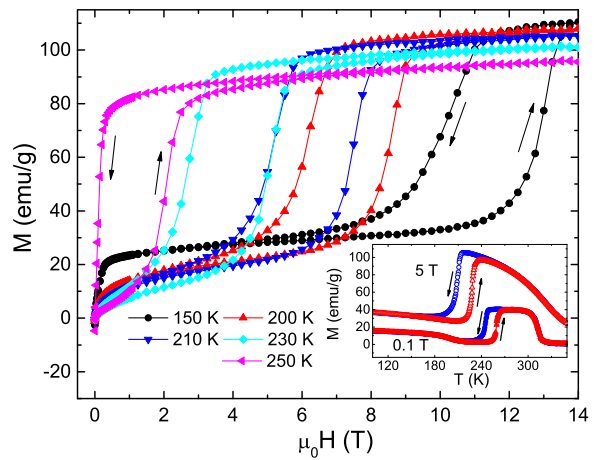}
  \caption{Isothermal magnetization data measured in magnetic fields up to 14~T for different temperatures below $T_A$. Each measurement was preceded by heating up the sample to the fully austenitic state and then cooling down to the completely martensitic state (100~K) before approaching the measurement temperature. The inset shows FC (circles) and FH (triangles) magnetization data for 0.1~T (closed symbols) and 5~T (open symbols).}\label{MHall14T_MT}
\end{center}
\end{figure}

The inset of Fig.\ \ref{MHall14T_MT} shows temperature-dependent magnetization curves, $M(T)$, of Ni$_{50}$Mn$_{35}$In$_{15}$ measured at 0.1 and 5~T, following FC (field cooled) and FH (field-cooled heating) protocols. The ferromagnetic transition of the austenitic phase takes place at $T_C\approx315$~K, and the martensitic transition from the austenitic to the martensitic phase is  on cooling observed at $T_M\approx246$~K and on heating around $T_A\approx 257$~K. The applied field of 5~T shifts the martensitic transition temperature by about -7~K, which indicates that the magnetic field stabilizes the austenitic phase. The $M(H)$ data in Fig.\ \ref{MHall14T_MT} indicate that it is possible to induce the reverse martensitic transition also by application of a magnetic field, at 150~K a field of 14~T is sufficient to observe a completed transition. For each $M(H)$ measurement the sample was cooled down from the 350~K (austenitic phase) to 100~K (complete martensitic phase) in zero field and then heated up to the target temperature of the experiment. By this protocol we assure that the sample is always in the same, fully martensitic, initial state and that we can exclude any influence of hysteresis effects on the results. At 150~K, we determined from the inflexion point in $M(H)$ a critical field of $H_A\approx13$~T for inducing the transition from the martensitic to the austenitic state. When the field is removed again, the sample is transformed back to the martensitic state. Note, for temperatures close to the transition at $T_A$ part of the sample remains in the austenitic phase after removing the magnetic field. We observe a hysteresis of $2-3$~T between the critical fields upon increasing and decreasing the magnetic field. Upon increasing the temperature toward $T_A$ the critical field, $H_A(T)$, decreases.

\begin{figure}[t!]
\begin{center}
  \includegraphics[width=0.9\linewidth]{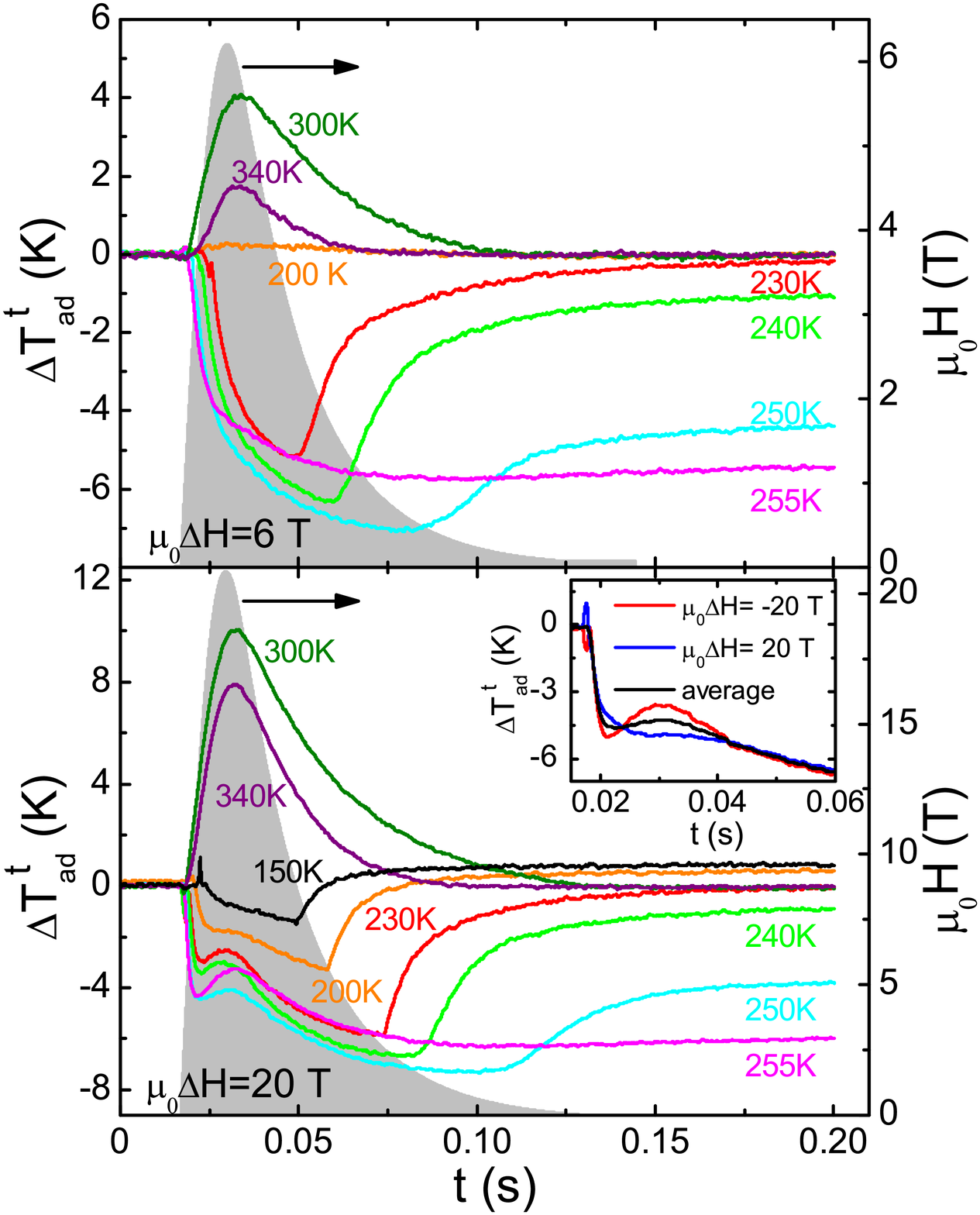}
  \caption{Time dependences of the adiabatic temperature change at 6~T (top) and 20~T (bottom), measured in pulsed magnetic fields following the protocol described in the text. The field profile is also indicated (right axis). The inset shows $\Delta T^{t}_{ad}(t)$ with pulsed fields of -20~T and 20~T at 240~K and the average of the two.}\label{raw}
\end{center}
\end{figure}

The MCE, in this case the adiabatic temperature change, $\Delta T_{ad}$, was measured by both direct and indirect methods. Figure \ref{raw} shows the time dependences $\Delta T^{t}_{ad}(t)$ for selected temperatures recorded during application of a pulsed field of 6 (top) and 20~T (bottom) together with the corresponding pulse profile. For pulsed-field measurements at temperatures below the martensitic first-order transition the sample was cooled down from 350 to 100~K in zero field and then heated up to the target temperature as in the case of the $M(H)$ measurements. Note, the little bumps observed at the beginning of the $\Delta T^{t}_{ad}(t)$ curve in the 20~T experiments are contributions from the magnetization change in the sample. The small open loops in the thermocouple wires act like magnetization pick-up coils and detect the metamagnetic transition of the sample. The inset of Fig.\ \ref{raw} exemplifies how a reversal of the field polarity affects the sign and magnitude of the bump, confirming that it is not related to the MCE response. To minimize this effect at all temperatures two curves taken on a positive and a negative field pulses have been averaged.

Around the Curie temperature in the austenitic phase, $T_C\approx315$~K, we observe a conventional MCE as expected (see the data at 300 and 340~K in Fig.\ \ref{raw}). As the field is applied the magnetic moments in the austenite fully align with the field and the magnetic entropy decreases. Since the measurement is performed adiabatically, the decrease in the magnetic entropy is compensated by an increase of lattice entropy leading to an increase in the temperature of the sample. Furthermore, the effect is reversible, i.e., after the magnetic pulse the sample reaches again the starting temperature. We point out that the position of the maximum of the pulse roughly coincides with that of the maximum in $\Delta T^{t}_{ad}(t)$.

On the other hand, the $\Delta T^{t}_{ad}(t)$ data at temperatures below the martensitic transition show a completely different behavior. When the field is applied, the structural transformation is induced and completed for all temperatures as shown in Fig.\ \ref{raw} (with the exception of the measurement at 200~K in the 6~T field range). This transformation involves a positive entropy change, which leads, in an adiabatically conducted experiment, to a negative temperature change (inverse MCE), $\Delta T_{ad}^{str}<0$. Furthermore, the structural transformation involves a change in the magnetic state of the sample. The magnetization increases involving a negative entropy change (conventional MCE) corresponding to a positive temperature change, $\Delta T_{ad}^{mag}>0$. Thereby, the structural contribution works against the magnetic contribution to the total MCE. As the structural contribution is dominant for temperatures close to the martensitic transition, an inverse MCE is observed, i.e., the sample cools down upon application of the magnetic field, $\Delta T_{ad}=\Delta T_{ad}^{str}+\Delta T_{ad}^{mag}<0$.\cite{Liu2012} Moreover, close to $T_A$ the inverse effect is irreversible and the sample does not recover the initial temperature after the magnetic field is removed and the maximum of the field profile does not coincide with the minimum in $\Delta T^{t}_{ad}(t)$.

For all temperatures (except for 200~K and $6$~T) the applied magnetic field is large enough to induce and observe a complete reverse martensitic transition (see Fig.\ \ref{MHall14T_MT}). Thus, a pure austenitic state is reached at the end of the field-up sweep. As we mentioned previously, a hysteresis of $2-3$~T is present between the induced transition in the up and down sweep. Henceforth, as the magnetic field is reduced again, we first observe a conventional MCE due to the change of the magnetic entropy in the ferromagnetic austenitic phase and the sample continues to cool down further.\cite{Liu2012,Khovaylo2010,Titov2012} As the field continues to decrease, the sample transforms back to the martensitic phase at the critical field of the field-induced martensitic transition leading to a negative change of the structural entropy. As a consequence the sample starts to heat up again, $\Delta T_{ad}^{str}>0$. Depending on the temperature the material either recovers the martensitic phase completely (see for instance curve at 230~K with field pulse of 6~T), or a mixed martensitic/austenitic  state is established (see the data at 240 and 250~K). In the latter case, only part of the sample transforms back and, therefore, a smaller change in temperature is observed in the down sweep compared to the up sweep. As a result, the final temperature of the sample is lower than the initial one. In particular, at 255~K the temperature of the sample after the field pulse remains below the starting temperature, consequently, the sample does not transform back to the martensite and no further heating is observed.

\begin{figure}[t!]
\begin{center}
  \includegraphics[width=0.9\linewidth]{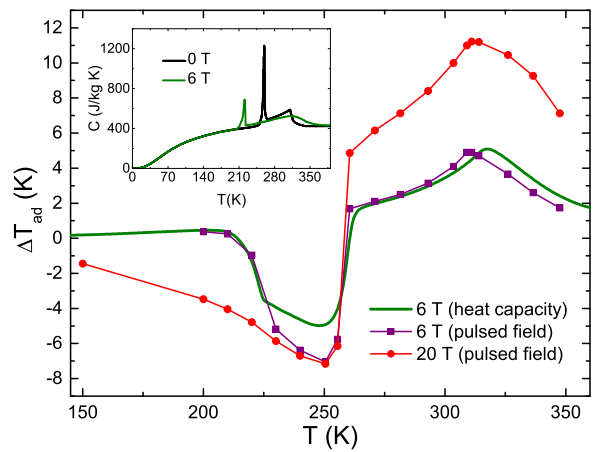}
  \caption{The adiabatic temperature change, $\Delta T_{ad}$, as a function of temperature obtained from heat-capacity data in a field of 6~T and magnetocaloric measurements in pulsed fields of 6 and 20~T. The inset displays the temperature dependence of the specific heat at 0 and 6~T.}\label{dTa}
\end{center}
\end{figure}

Additionally to the MCE experiments in pulsed magnetic fields, we carried out specific-heat measurements in zero field and in an applied field of 6~T. For the measurement in magnetic field the sample was cooled in zero field from the austenitic phase down to 2~K and the specific heat was recorded on heating from 2 to 380~K. At the first-order martensitic transition we carefully analyzed the slope of the temperature profile of a long heat pulse to obtain the heat capacity.\cite{J.C.Lashley2003} This procedure avoids an underestimation of the heat capacity due to the latent heat at the first-order phase transition.\cite{J.C.Lashley2003,Hardy2009} The specific-heat data are presented in the inset of Fig.\ \ref{dTa}. From these results we calculated the temperature dependence of the total entropy for 0 and 6~T and subsequently the adiabatic temperature change, $\Delta T_{ad}$. In the calculations we have used the thermodynamic relations $S(T)_{H_I} \approx \int_{0}^{T}{[C(T)_{H_I}/T]}dT$, $S(T)_{H_F} \approx \int_{0}^{T}{[C(T)_{H_F}/T]}dT$,  and  $\Delta T_\text{ad}(T)_{\Delta H} \approx [T(S)_{H_F}-T(S)_{H_I}]_S$ where $H_I$ and $H_F$ indicate the initial and final magnetic field values, respectively.\cite{Pecharsky_JOAP_1999a} $\Delta T_{ad}(T)$ obtained from the heat-capacity data is presented in Fig.\ \ref{dTa} together with the results of the direct measurements. We note that due to the finite response time of the thermocouple and the pick-up of a magnetic signal due to the change in magnetization at the metamagnetic transition, it is not possible to trace the time (field) dependence of $\Delta T^{t}_{ad}$ precisely, only the maximum/minimum in $\Delta T^{t}_{ad}(t)$ is well defined and those values are plotted as $\Delta T_{ad}$ in Fig.\ \ref{dTa}. We find, as expected, an inverse MCE in the region below the martensitic transition and a conventional MCE around the ferromagnetic transition of the austenitic phase, i.e.,\ in an adiabatic environment the sample cools down and warms up, respectively, upon increasing the magnetic field. An important feature, which we want to point out is that the MCE just below the martensitic transition saturates already in small magnetic fields. $\Delta T_{ad}$ reaches a value of around -7~K at 6~T which does not change up to 20~T anymore. With a field change of 20~T, $\Delta T_{ad}$ only displays a broader peak, since the reverse martensitic transition can be induced also at lower temperatures. Our findings confirm that the inverse MCE due to the first-order martensitic transition is limited by the latent heat of the transition.\cite{Emre2013} We note that the martensitic transition can be induced by a field of 20~T even at 4~K.\cite{Nayak2014} With increasing temperature this critical field decreases and for 150~K the transition is already completed at 14~T, as indicated in Fig.\ \ref{MHall14T_MT}. However, the MCE measured at this temperature is only 20\% of the maximum value at 250~K. This behavior is related to the decrease in the entropy change at the structural transition with decreasing temperature.\cite{SalazarMejia}

Around the second-order ferromagnetic transition in the austenitic phase ($T_C\approx315$~K) we find a good agreement between the results obtained from the heat capacity and from the MCE measurements in pulsed fields, but there are distinct discrepancies below the first-order martensitic transition. At temperatures just below the martensitic transition the minimum in the time-depended temperature change, $\Delta T^{t}_{ad}(t)$, is observed for a magnetic field beyond the pulse maximum on the down sweep. This maximum value in $|\Delta T^{t}_{ad}(t)|$ corresponds, therefore, not only to the MCE as it is generally defined. As we have discussed before, at these temperatures an inverse MCE is observed, i.e., the sample cools down due to a dominant structural contribution ($\Delta T_{ad}^{str}<0$ and $\Delta T_{ad}^{mag}>0$). This is the value of $\Delta T_{ad}$ obtained by the specific-heat experiments in static fields, but in case of the pulsed-field experiments decreasing the magnetic field leads to a further cooling of the sample, since $\Delta T_{ad}^{mag}<0$ and, initially, $\Delta T_{ad}^{str}\approx0$. Only once the field-induced structural transition takes place upon further lowering the magnetic field, $\Delta T_{ad}^{str}$ provides a positive contribution to $\Delta T^t_{ad}(t)$.

\begin{figure}[t!]
\begin{center}
  \includegraphics[width=0.9\linewidth]{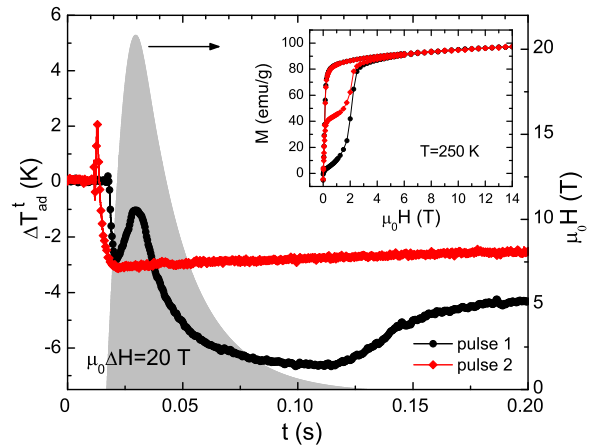}
  \caption{Time dependence of $\Delta T^{t}_{ad}$ measured at 250~K for a magnetic-field pulse of 20~T. The initial curve (black circles) was measured following the standard protocol as discussed in text) and the second curve (red diamonds) was taken 1 hour after the first pulse. The field profile is also shown (right axis). The inset shows the corresponding $M(H)$ curves following the same protocol.}\label{M(H)_repeat}
\end{center}
\end{figure}

The observed irreversibility in the MCE is expected in shape-memory Heusler materials.\cite{Khovaylo2010,Emre2013} It originates from the thermal/magnetic hysteresis at the first-order martensitic phase transition. Therefore, field cycling or repeated measurements at the same temperature without considering the magnetic and thermal history of the sample lead to different results in the MCE. To study the influence of the irreversibility of the transition on the MCE for temperatures close to $T_A$, we have measured $M$ and $\Delta T^t_{ad}$ in two subsequent magnetic field sweeps. The second pulse was applied after waiting 1 hour after the first pulse, which was performed following our standard protocol as mentioned above. Figure \ref{M(H)_repeat} displays the results at 250~K for $\Delta T^{t}_{ad}(t)$ and $M(H)$ (inset). The $M(H)$ data indicate that pulse 1 induces the reverse martensitic transition and that the sample stays in the austenitic phase. After 1 hour, only a small part of the sample has transformed back to the martensitic phase. Thus, the entropy change due to the second pulse is much smaller and the maximum temperature change of the sample is less than $1/2$ of that observed in the first pulse, $|\Delta T_{ad}|\approx6$~K. This observation is consistent with the $M(H)$ data obtained during the second pulse, i.e., only part of the sample is in the martensitic state and can still transform to the austenitic phase. Furthermore, $\Delta T^{t}_{ad}(t)$ stays almost constant after the initial decrease; the temperature relaxes only very slowly towards the starting temperature. This behavior hampers the use of shape-memory alloys in possible applications since to have a maximum cooling effect the sample needs to be brought to the initial conditions before each cooling cycle.

In summary, we have investigated the MCE in the shape-memory Heusler alloy Ni$_{50}$Mn$_{35}$In$_{15}$ by direct magnetocaloric measurements in pulsed magnetic fields and by heat-capacity experiments in constant fields. The conventional MCE around the Curie temperature in the austenitic phase exhibits a strong magnetic-field dependence, for a field change of 20~T we find a maximum $\Delta T_{ad}\approx11$~K. In this region of the phase diagram we do not find any hysteresis or irreversible behavior in the magnetocaloric experiments in pulsed-magnetic fields. Furthermore, $\Delta T_{ad}$ is in a good agreement with the data from heat capacity. This changes below the martensitic phase transition. Here, we observe an inverse MCE, which originates from structural and magnetic contributions. In the direct magnetocaloric experiments we find a saturating MCE with a maximum negative temperature change of $\Delta T_{ad}= -7$~K. This value is significantly larger than the results of the specific-heat analysis. The cause for this lies in the hysteresis observed at the martensitic phase transition which leads to a strongly irreversible time dependence of  $\Delta T^t_{ad}$ below the martensitic phase transition. This irreversible behavior has to be carefully taken into consideration for using shape-memory Heusler alloys in magnetocaloric-cooling applications, especially, since the length of the magnetic field pulse is comparable to the inverse of useful operation frequencies of cooling devices.

We acknowledge the support of the HLD at HZDR, member of the European Magnetic Field Laboratory (EMFL).
This work was financially supported by the ERC Advanced Grant (291472) "Idea
Heusler."

%

\end{document}